\documentclass[twocolumn,showpacs,aps]{revtex4}
\usepackage{amsmath,amssymb,bm}
\usepackage{graphicx,color}
\usepackage{hyperref}
\usepackage{epstopdf}

\newcommand{\Opm}{\hat O_{\mbox{\tiny $\pm$}}}

\newcommand{\tS}{\mbox{\tiny S}}
\newcommand{\B}{\mbox{\tiny B}}
\newcommand{\SB}{\mbox{\tiny SB}}
\newcommand{\T}{\mbox{\tiny T}}
\newcommand{\w}{\omega}
\newcommand{\ti}{\Tilde}
\newcommand{\wti}{\widetilde}
\newcommand{\dg}{\dagger}
\newcommand{\la}{\langle}
\newcommand{\ra}{\rangle}
\newcommand{\La}{\big\la}
\newcommand{\Ra}{\big\ra}

\newcommand{\nl}{\nonumber \\}
\newcommand{\Sec}[1]{Sec.\;\ref{#1}}
\newcommand{\App}[1]{Appendix\;\ref{#1}}
\newcommand{\be}{\begin{equation}}
\newcommand{\ee}{\end{equation}}
\newcommand{\bsube}{\begin{subequations}}
\newcommand{\esube}{\end{subequations}}
\newcommand{\Eq}[1]{Eq.\,(\ref{#1})}
\newcommand{\Eqs}[1]{Eqs.\,(\ref{#1})}
\newcommand{\Fig}[1]{Fig.\,\ref{#1}}

\begin{document}

\title{Nonequilibirum noise spectrum and Coulomb--blockade--assisted
Rabi interference \\ in a double-dot Aharonov-Bohm interferometer}

\author{Jinshuang Jin} \email{jsjin@hznu.edu.cn}
\affiliation{Department of Physics, Hangzhou Normal University,
 Hangzhou, Zhejiang 311121, China}

\begin{abstract}

We investigate the charge--states coherence
underlying the nonequilibirum transport through
a spinless double-dot Aharonov-Bohm
(AB) interferometer.
Both the current noise spectrum and real-time dynamics
are evaluated with the well--established
dissipaton--equation--of--motion method.
The resulted spectrums show the characteristic peaks and dips,
arising from coherent Rabi oscillation dynamics,
with the \emph{environment--assisted} indirect
inter-dot tunnel coupling mechanism.
The observed spectroscopic features are in a quantitative
agreement to the real-time dynamics of the reduced density
matrix off–diagonal element between two charge states.
As the aforementioned mechanism,
these characteristics of coherence are
very sensitive to the AB phase.
While this is generally true
for cross-correlation spectrum,
the total circuit noise spectrum that is experimentally
more accessible shows remarkably rich interplay
between various mechanisms.
The most important finding of this work is
the existence of Coulomb--blockade--assisted Rabi interference,
with very distinct signatures arising from
the interplay between the AB interferometer
and the interdot Coulomb interaction
induced Fano resonance.

\end{abstract}


\maketitle

\section{Introduction}

 Quantum transport through a parallel double-dot
embedded in Aharonov-Bohm (AB) interferometer
has been attracted much attention, in both experiments \cite{Sch97417,Hol01256802,Sig06036804,Sig04066802,Kub06205310,Hat11076801}
and theories \cite{Liu16045403,Kan04117,Szt07386224,Kon013855,%
Kon02045316,Li09521,Tok07113,Bed14235411,Rep16165425}.
This is a promising candidate for
solid state qubit devices \cite{Hol01256802,Bur992070}.
The observed transport current shows the AB oscillation
as a function of the magnetic flux \cite{Hol01256802,Sig06036804}.
Physically, the observed features
exhibit a rich interplay between
Coulomb interaction \cite{Kon013855,Kon02045316,%
 Li09521,Tok07113,Bed14235411,Szt07386224},
interdot tunneling \cite{Kan04117,Kub06205310,%
 Szt07386224,Hat11076801,Liu16045403},
inelastic electron cotunneling
processes \cite{Sig06036804,Rep16165425},
and so on.
Recent interests include the dynamical effect
of charge-state coherence
on transient transport in double--dots AB interferometer
\cite{Har13235426,Bed13045418,Tu12115453,Tu11115318,Bed12155324,Jin18043043}.
It is found that in the absence of Coulomb interaction,
the charge qubit states experience the phase localization
at values of $m\pi$ \cite{Tu11115318,Bed12155324}.
On the other hand, the interdot Coulomb interaction
opens Coulomb--assisted channels \cite{Jin18043043},
which interfere with
the single-electron impurity channels.
The relative phase is no longer localized but continuously
controllable with the magnetic flux \cite{Jin18043043}.
 Moreover, an almost decoherence--free charge qubit state
emerges at the current--voltage turnover position \cite{Jin18043043},
where differential conductance is zero.
This is the  double-dot electron pair tunneling resonance
condition \cite{Lei09156803},
where the chemical potential of one
electrode is at the center between
a single-electron impurity channel and the related Coulomb-assisted
channel.

Comparing to the average current characteristics,
the shot noise spectrum offers much detailed insights
in nonequilibirum quantum transport
\cite{Her927061,Bla001,Imr02,Naz03}.
The zero--frequency noise describes mainly the
statistical behavior of the steady state.
It has been used to measure the effective carrier charge that can be
either fraction \cite{Pic97162,Rez99238,Bid09236802} or integer
\cite{Koz003398,Lef03067002}.
As anticipated, shot noise spectrum is more sensitive than current to
quantum interference, which is particularly prominent
in such as the double--dot AB interferometer in the Kondo regime
\cite{Zha06085106,Fan07205312,Bre11155305}.
The evaluated shot noise oscillation behavior in Ref.\ \onlinecite{Zha06085106}
well reproduced that of experiment \cite{Hol01256802}.

 Recent progress in on-chip detection allows
the high--precision measurement
of quantum current--current correlator spectrum
in the full frequency range
\cite{Agu001986,Deb03203,Ona06176601,Zak07236803,Bas10166801,Bas12046802,Del18041412}.
Nonequilibrium quantum noise spectrum
contains rich correlated dynamics information
\cite{Ent07193308,Li05066803,Bar06017405,Gab08026601,Wab09016802,
Eng04136602,Jin11053704,Jin13025044,Rot09075307,Wan13035129}.
 It is neither asymmetric nor detailed--balance related,
with respect to the sign of frequency.
However, most of theoretical studies
were carried out for symmetrized noise spectrum,
based on the MacDonald's formula \cite{Mac62}.
This is a Markovian and quasiclassical treatment.
Nevertheless, the evaluated  spectrum
does show the oscillation dynamics induced by
interdot transfer couplings \cite{Don08033532}.
It had also identified
a pronounced super/sub-Poissonian statistics
at spin-spin interaction energies \cite{Ziv11115304}.

 In this work, we investigate the nonequilibrium
quantum current noise spectrum
for the nondegenerate double-dot AB (DD-AB) interferometer systems.
We do not include the Kondo effect
that is important in the low temperature regime.
Both lead-specified and total circuit current noise spectrums are
 accurately evaluated
based on the well--established dissipaton--equation--of--motion (DEOM) theory \cite{Yan14054105,Yan16110306,Jin15234108}.
It is well-known that the total noise spectrum is the consequence of the
lead-specified current-current interference.
Our main results but not limited are as follows.
(\emph{i}) The current noise spectrums
show rich characteristics in relation to
the coherent Rabi oscillation.
This is in a good agreement to
the real-time dynamics of the reduced density matrix
off--diagonal element between two charge states;
(\emph{ii}) The observed Rabi characteristics arise from
the environment-assisted indirect interdot
coupling;
(\emph{iii}) In the absence of the interdot Coulomb interaction,
the total circuit noise spectrum for pristine double-dot
(AB phase $\phi=0$) only shows the non-Markovian quasi-step
features. There are no Rabi signal, which exists in the
lead-specified auto-correlation and cross-correlation current noise spectra.
 The disappearance of the Rabi signal in the total circuit noise
 is due to the lead-specified current destructive interference;
(\emph{iv}) However, the Rabi dips do appear when AB phase is nonzero
and reach maximum at $\phi=\pi$;
(\emph{v}) Remarkably, in the presence of the
strong interdot Coulomb interaction, the Rabi dips
in the total circuit noise spectrum become Fano profiles.
This is the so-called Coulomb–blockade–assisted (CB--assisted) Rabi
interference phenomenon.
Its signature in the circuit noise develops from the peak at
AB phase $\phi=0$ to dip at $\phi=\pi$,
and Fano line shape in between.
The remainder of this paper is organized as follows.
In \Sec{thMet}, we present the model
and the quantities of interest,
with their evaluations via the DEOM approach
being detailed in \App{thDEOM}.
In \Sec{thnum}, we demonstrate the numerically
accurate results of the current noise spectra,
including the auto-correlation, cross-correlation and the total
circuit ones together with the real-time dynamics of reduced density matrix.
First, we consider the noninteracting scenarios in \Sec{thnonee}.
We then exploit the effect of interdot Coulomb interaction
and CB--assisted Rabi interference in \Sec{thCou} and \Sec{thCouRabi}, respectively.
In line with the experimental realization
of AB interferometer \cite{Hat11076801},
we discuss the effect of the indirect interdot
tunneling strength, by introducing the indirect coupling parameter \cite{Kub06205310},
on the noise spectra and the real-time
dynamics of reduced density matrix in \App{thlam}.
Finally, we conclude this work with \Sec{thsum}.
%

\section{Methodology}
\label{thMet}

 Consider a parallel double-dot
embedded in an AB interferometer
and coupled to the electron reservoirs.
The total Hamiltonian is composed of the three parts, $H_{\T}=H_{\tS}+H_{\B}+H_{\SB}$.
The double-dot system Hamiltonian is modeled by
\be\label{Hs}
 H_{\tS} =\sum_{u=1,2}\varepsilon_u \hat a^{\dagger}_{u}\hat a_{u}
 +U \hat n_1\hat n_2.
\ee
Here, $\hat a_{u}$ ($\hat a^{\dagger}_{u}$ )
denotes the annihilation (creation)
operator of the electron in
dot-$u$ with the energy level $\varepsilon_u$,
$\hat n_u=\hat a^{\dg}_{u}\hat a_{u}$ is the electron occupation,
and $U$ is the interdot, capacitive Coulomb interaction strength.
The involved charge states in the double dots are $|0\ra=|00\ra$, $|1\ra=|10\ra$, $|2\ra=|01\ra$,
and $|d\ra\equiv|11\ra$, i.e., the empty, the dot-$1$
 occupied, the dot-$2$ occupied, and double--dots--occupancy states,
respectively. The two  single-electron charge states of
$|1\ra=|10\ra$ and $|2\ra=|01\ra$
can be served as a charge qubit \cite{Agu04206601,Luo07085325,Kie07206602,Tu08235311}.
The quantum coherence properties of
the double-dot states are described by
the reduced system density matrix,
$\rho(t)\equiv {\rm tr}_{\B}\rho_{\rm tot}(t)$, i.e.,
 the partial trace of the total density operator
$\rho_{\rm tot}$ over the electrode bath degrees of freedom.

The environment bath Hamiltonian of the two electron reservoirs
is given by
\be\label{HB}
 H_{\B} = \sum_{\alpha k}(\epsilon_{\alpha k}+\mu_{\alpha})
  \hat c^{\dg}_{\alpha k}\hat c_{\alpha k}.
 \ee
Here,  $\hat c^{\dg}_{\alpha k}$ ($\hat c_{\alpha k}$)
denotes the creation (annihilation) operator of the electron
with momentum $k$ and energy $\epsilon_{\alpha k}$,
in the left ($\alpha={\rm L}$) or right ($\alpha={\rm R}$) reservoir,
under the applied bias voltage potential,
$eV=\mu_{\rm L}-\mu_{\rm R}$.
 The system--bath coupling assumes
the standard tunneling form, which
in the presence of an AB interferometer
reads
\be\label{Hsb}
  H_{\SB}=\sum_{\alpha u k} \left( e^{i\phi_{\alpha u}}t_{\alpha u k}
  \hat a^{\dg}_{u}\hat c_{\alpha k}  +{\rm H.c.}\right).
\ee
The AB phases via threading a magnetic flux $\Phi$
satisfy $\phi_{{\rm L}1}-\phi_{{\rm L}2}
+\phi_{{\rm R}2}-\phi_{{\rm R}1}=\phi\equiv
 2\pi\Phi/\Phi_{0}$, where $\Phi_{0}$ is the flux quantum.
Without loss of generality,
we adopt
$ \phi_{{\rm L}1}=-\phi_{{\rm L}2}
 =\phi_{{\rm R}2}=-\phi_{{\rm R}1}=\phi/4$,
due to the gauge invariant \cite{Tu11115318,Bed13045418}.
To complete the description,
one requires the reservoir hybridization spectral function,
$J_{\alpha uv}(\omega)
\equiv \pi e^{i(\phi_{\alpha v}-\phi_{\alpha u})}
 \sum_k  t^\ast_{\alpha u k}t_{\alpha v k}\delta(\omega-\epsilon_{\alpha k})$.
For simplicity, we set it a Lorentzian-type form \cite{Mac06085324,Jin08234703,Tu08235311},
\be\label{J_Drude}
 J_{\alpha uv}(\omega)
 =\frac{\Gamma_{\alpha uv}W^2}{\omega^2+W^2},
\ee
with $\Gamma_{\alpha uu}=\Gamma_\alpha$ and
\be\label{Gam}
\begin{split}
 \Gamma_{{\rm L}12}&=\Gamma^\ast_{{\rm L}21}
   =\lambda_{\rm L} \Gamma_{\rm L} e^{-i\phi/2},
\\
 \Gamma_{{\rm R}12}&=\Gamma^\ast_{{\rm R}21}
 =\lambda_{\rm R} \Gamma_{\rm R} e^{i\phi/2}.
\end{split}
\ee
Here, $0 \leq |\lambda_{\alpha}| \leq 1$,
 is the indirect interdot coupling
parameter \cite{Kub06205310,Hat11076801}.
Moreover, the AB interferometer takes effect
only when $\lambda_{\alpha}\neq 0$.
Throughout this work, we adopt the unit of $e=\hbar=1$,
for the electron charge and the Planck constant.

 The current operator for the electron
transfer from $\alpha$-reservoir to
impurity system is
$\hat I_{\alpha}\equiv -\dot{\hat N}_{\alpha}
=-i[\hat N_{\alpha},H_{\T}]$.
Here, $\hat N_{\alpha}=\sum_{k}\hat c^{\dg}_{\alpha k}\hat c_{\alpha k}$
is the electron number in
the $\alpha$-reservoir.
Let $I^{\rm st}_{\alpha}$ be the steady--state current ($I^{\rm st}_{L}=-I^{\rm st}_{R}
\equiv \bar I$)
and $\delta{\hat I}_\alpha(t)\equiv{\hat I}_\alpha(t)-I^{\rm st}_{\alpha}$.
The fluctuating current correlation
function is then
\be
  \La \delta{\hat I}_\alpha(t)\delta{\hat I}_{\alpha'}(0)\Ra
=\La {\hat I}_\alpha(t){\hat I}_{\alpha'}(0)\Ra
  -I^{\rm st}_{\alpha}I^{\rm st}_{\alpha'}.
\ee
The susceptibility of the shot noise fluctuation is related
to the half-Fourier transform of
\be\label{Cw_alp}
 C_{\alpha\alpha'}(\omega)\equiv \int_0^{\infty} \!dt\,
  e^{i\omega t} \La \delta{\hat I}_\alpha(t)\delta{\hat I}_{\alpha'}(0)\Ra\,.
\ee
The lead-specified current noise spectrum via
the full Fourier transformation is then
\be\label{Sw_alp}
  S_{\alpha\alpha'}(\omega)=C_{\alpha \alpha'}(\omega)
    +C^{\ast}_{\alpha' \alpha}(\omega).
\ee
Its values at positive ($\omega>0$) and negative ($\omega<0$)
frequencies correspond to energy
absorption and emission processes,
respectively \cite{Eng04136602,Agu001986,Jin15234108}.
 It is worth noticing that in the equilibrium case,
the absorptive and emissive components are related each other via
the detailed--balance relation or the equivalent
fluctuation--dissipation theorem.
However, they are generally \emph{independent}
in a nonequilibrium scenario.
The widely studied symmetrized noise spectrum
corresponds to
$S^{{\rm sym}}_{\alpha\alpha'}(\omega)
=S_{\alpha\alpha'}(\omega)
+S_{\alpha'\alpha}(-\omega)$.
This can not distinct the nonequilibrium absorption
versus emission processes \cite{Rot09075307}.

 Moreover, the net circuit current in experiments is given by
$ I(t) = a I_{\rm L}(t)-b I_{\rm R}(t)$,
with the junction capacitance parameters,
$a, b\geq 0$, satisfying $a + b = 1$ \cite{Bla001,Eng04136602,Del18041412}.
The circuit current noise spectrum is then
$ S(\omega) = a^2S_\text{LL}(\omega)+b^2S_\text{RR}(\omega)
-2ab\,{\rm Re}[S_\text{LR}(\omega)]$,
and for the symmetric structure of  $a=b=1/2$,
it reads
\be\label{Sw}
 S(\omega) = \frac{1}{4}\big\{S_\text{LL}(\omega)
 +S_\text{RR}(\omega)-2\,{\rm Re}[S_\text{LR}(\omega)]\big\}.
\ee

Numerical evaluations will be carried out
via the fermionic DEOM method \cite{Yan14054105,Jin15234108,Yan16110306};
see \App{thDEOM} for the details.
This is a quasi-particle extension to the well--established
hierarchical--equations-of-motion
formalism \cite{Jin08234703}.
As an efficient and universal
method for strongly correlated
quantum impurity systems \cite{Zhe121129,Li12266403,%
Zhe13086601,Hou15104112,Ye16608},
DEOM converges rapidly and uniformly to the exact results,
with increasing the truncated tier level, $L=n_{\rm trun}$.
The minimal truncation tier $L$ required to achieve convergence is closely dependent on the
configurations of system as well as bath.
In practice, the convergence with respect to $L$
is tested case by case.
For the parameters exemplified in the
following numerical calculations,
the DEOM evaluations effectively
converge at $L=3$ tier level.


\section{Numerical results}
\label{thnum}
To clarify the physical picture,
we consider the symmetrical--leads situation, where
$\mu_{\rm L}=-\mu_{\rm R}=V/2$,
$\Gamma_\text{L}=\Gamma_\text{R}=\Gamma$
and $\lambda_\text{L}=\lambda_\text{R}=\lambda$.
Moreover, we focus on
the sequential-dominated tunneling regime,
with $\mu_{\rm L}>\varepsilon_1,\varepsilon_2>\mu_{\rm R}$ and $k_BT \gtrsim\Gamma$,
under the environment-assisted indirect interdot tunnel coupling with $\lambda=1$.
The effect of the indirect coupling parameter $\lambda$ on the current noise
spectra and the corresponding coherent charge dynamics of the charge states will be given in
the \App{thlam}.
Set the wide bandwidth value of $W=10$\,meV
for the electron reservoirs.
The details of other parameters are given
in the captions of the figures.
It is worth noting that the characteristics on the auto-correlation
noise spectra, $S_{\rm LL}(\omega)$ and $S_{\rm RR}(\omega)$, are
quite similar; thus only the
results of $S_{\rm LL}(\omega)$,
together with ${\rm Re}[S_{\rm LR}(\omega)]$,
are explicitly reported below.

To elucidate the underlying mechanisms,
the reduced system density matrix charge states dynamics
will also be reported.
Note that the degenerate
DD--AB systems ($\Delta\varepsilon=0$) were studied before
\cite{Jin18043043,Tu11115318,Bed12155324}.
The resulted density matrix evolves smoothly without oscillations.
For noninteracting case, the indirect interdot coupling would lead to
 a charge qubit phase localization \cite{Tu11115318,Bed12155324}.
However, for strong interacting case, the phase is
 continuously tuned by AB flux \cite{Jin18043043}.
Moreover, the resulted charge qubit is an almost coherent pure state \cite{Jin18043043}.
The present paper focuses on the nondegenerate case ($\Delta\varepsilon\neq0$),
elucidating in particular the spectroscopic signature
of the two charge states with finite energy--splitting.

\subsection{Noninteracting scenarios}
\label{thnonee}

\begin{figure}
\includegraphics[width=1.0\columnwidth]{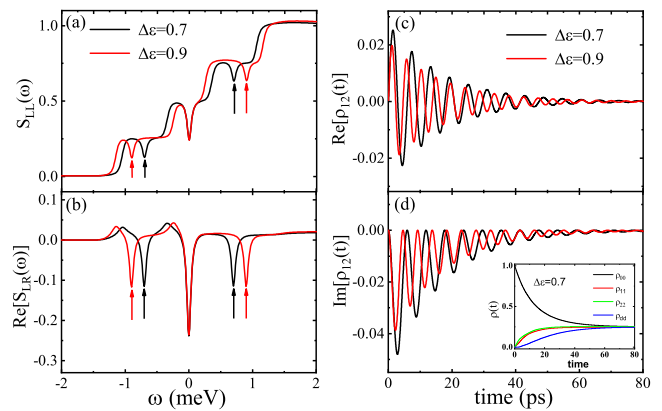}
\caption{ (Color online)
The accurate results for the current noise spectra (in $2\bar I$) and
the real-time dynamics of the charge states in the double-dot,
for noninteracting $U=0$ and AB phase $\phi=0$ with different
 energy splitting $\Delta\varepsilon$. Here, we set $\varepsilon_1=\Delta\varepsilon/2$,
 and $\varepsilon_2=-\Delta\varepsilon/2$.
(a) The auto-correlation noise spectrum of the left lead, $S_\text{LL}(\omega)$.
 (b) The cross-correlation noise spectrum of the left-right leads,
 ${\rm Re}[S_\text{LR}(\omega)$].
(c) and (d) The time evolution of the off-diagonal density matrix $\rho_{12}(t)$ with
the real and imaginary parts, respectively.
 The inset in (d) is the dissipative dynamics of the
probabilities of the four charge states in the double-dot.
The other parameters are (in ${\rm meV}$) $eV=1.4$,
$\Gamma=0.01$ and $k_{\B} T=0.02$. }
\label{fig1}
\end{figure}

 This subsection is concerned with the noninteracting scenario;
i.e., $U=0$ in the Hamiltonian \Eq{Hs}.
To single out different effects,
we consider first the pristine double--dots system, in the absence
of magnetic flux ($\phi=0$).
 Figure \ref{fig1} displays the numerically
accurate results on the noise spectra
and the real-time charge--states dynamics
in the pristine system.
As seen in \Fig{fig1}(a) and (b),
the noise spectrums of both auto and cross correlations
show dips at the energy splitting,
i.e., $|\omega|=\Delta\varepsilon$ with $\Delta\varepsilon=\varepsilon_1-\varepsilon_2$.
This feature is similar to
that of the intrinsic coherent Rabi oscillation induced by the
direct interdot coherent coupling in both serial and parallel coherent
coupled double-dot \cite{Luo07085325,Don08033532}.
Indeed, we find the coherent Rabi oscillation dynamics between the two
non-degenerate charge states, $|1\ra$ and $|2\ra$
in the double-dot. This is concerned with
the reduced density matrix element of $\rho_{12}(t)$,
as plotted in \Fig{fig1}(c) and (d).
In contrast, the population dynamics of the two charge states,
as reported in the inset in \Fig{fig1}(d),
show no oscillations at all.
While the Rabi oscillation frequency occurs at the energy splitting,
$\omega=\pm\Delta\varepsilon$,
its manifestation as a quantum interference transport
goes also by the
environment-assisted indirect interdot tunnel coupling
($\Gamma_{\alpha 12}\neq 0$).
It will be further demonstrated in the \App{thlam}.

 It is also noticed the Lorentzian-like (Markovian)
dip at $\omega=0$, in both the auto-- and cross--correlation
noise spectra. This feature arises from
the Pauli exclusion principle and
is dictated by long--time dynamics \cite{Jin15234108}.
Moveover, the transport induced transitions
show in $S_{\alpha\alpha}(\omega)$
the quasi-steps rising around $\omega=\pm|\varepsilon_u-\mu_\alpha|$,
whereas in ${\rm Re}[S_{\rm LR}(\omega)]$,
they are the non-Markovian
peaks at $\omega=-|\varepsilon_u-\mu_\alpha|$.
These features are also consistent with
our previous study on a single-impurity Anderson system \cite{Jin15234108}.
They affect distinctly the aforesaid dips
at the Rabi frequency, $\omega=\pm\Delta\varepsilon$,
which appear more remarkably in the cross-correlation noise
spectrum than the auto--correlation ones.
The observations here also agree well with
the Rabi dips nature of the indirect interdot tunnel coupling.

\begin{figure}
\includegraphics[width=1.0\columnwidth]{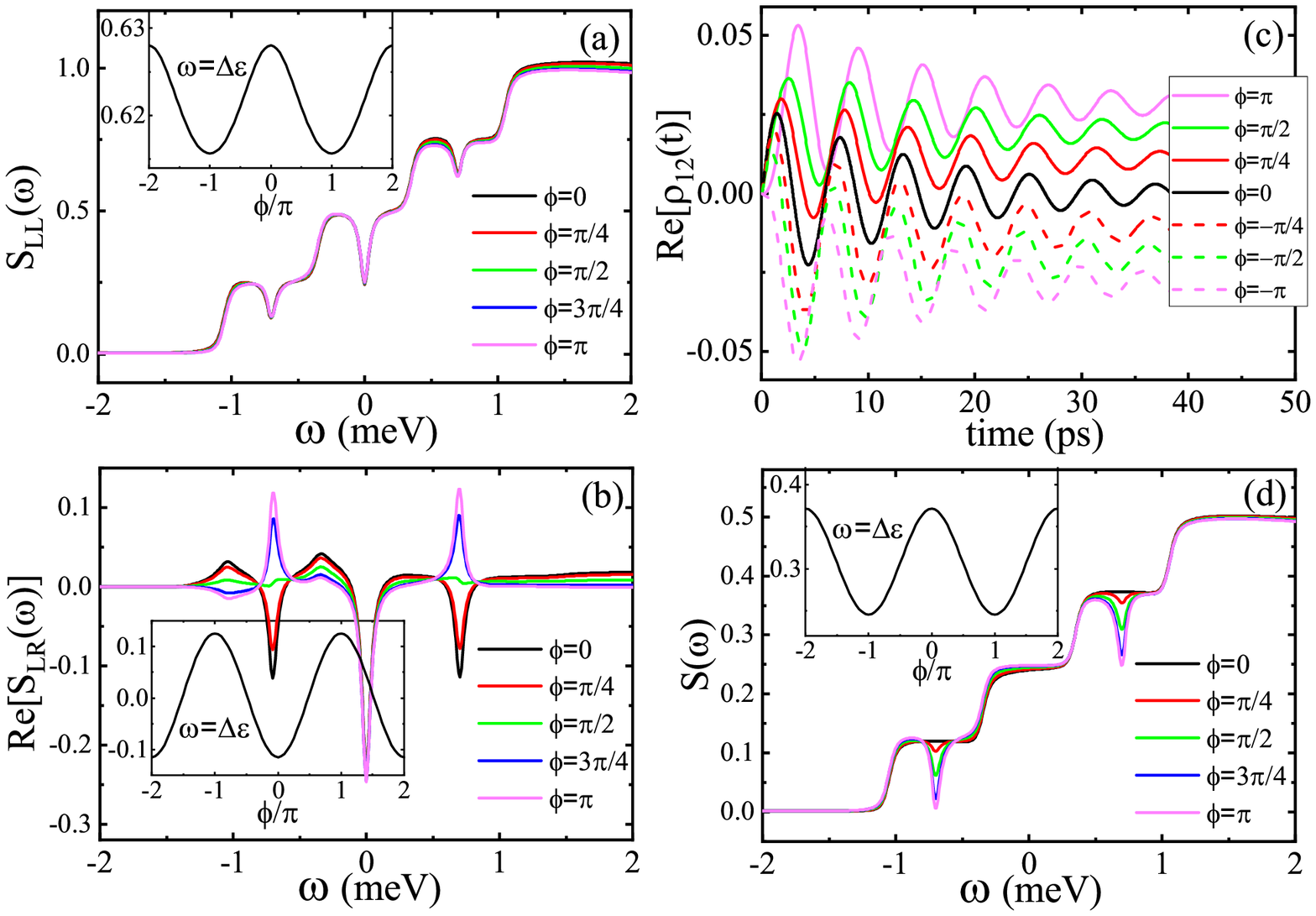}
\caption{ (Color online)
The accurate results for the current noise spectra (in $2\bar I$)
and the real-time dynamics of
the reduced density matrix off-diagonal element,
with different AB phase $\phi$ for noninteracting $U=0$
and energy splitting $\Delta\varepsilon=0.7{\rm meV}$.
(a) The auto-correlation noise spectrum of the left lead,
with the oscillation signal in the absorption part as a function of
the AB phase in the inset.
 (b) The cross-correlation noise spectrum,
 with the oscillation signal in the absorption part as a function of
the AB phase in the inset.
(c) The time evolution of  ${\rm Re}[\rho_{12}(t)]$.
 (d) The circuit noise spectrum,
 with the Rabi signal in the absorption part as a function of
the AB phase in the inset.
 The other parameters are the same as in \Fig{fig1}.
}
\label{fig2}
\end{figure}

 Explore now the effect of AB phase $\phi$
by treading the magnetic flux $\Phi$,
exemplified with $U=0$ and $\Delta\varepsilon=0.7$\,meV.
Interestingly, the auto-correlation noise
spectrum is weakly modified by the flux,
as displayed in \Fig{fig2}(a).
The inset highlights that $S_{\rm LL}(\w=\Delta\varepsilon)$
is a periodic function of the AB phase, but
of a small amplitude change.

In contrast, the cross-correlation noise spectrum,
as shown in \Fig{fig2}(b), is
very sensitive to the AB phase.
In particular, ${\rm Re}[S_\text{LR}(\w)]$,
at the characteristic oscillation frequency, $\omega=\pm\Delta\varepsilon$,
changes from a valley at $\phi=0$ to a peak feature at $\phi=\pm\pi$.
It satisfies ${\rm Re}[S_\text{LR}(\Delta\varepsilon)]
\approx{\rm Re}[S_\text{LR}(-\Delta\varepsilon)]\propto - \cos\phi$;
see the inset of \Fig{fig2}(b).
In parallel, we present $\rho_{12}(t)$
in \Fig{fig2}(c), at different values of the AB phase.
This reduced system density matrix element
describes coherence
between the two specified non-degenerate charge states,
$|1\ra$ and $|2\ra$, in the double-dot.
The dynamical phase, which could be represented with
${\rm Re}\rho_{12}(t=2\pi/\Delta\varepsilon)$,
is tuned by the AB phase.
The observed time-domain Rabi oscillation dynamics,
which manifests the periodic behavior with the AB phase
and does reflect in
${\rm Re}[S_\text{LR}(\w=\pm\Delta\varepsilon)]
\propto -\cos \phi$.
This is purely the quantum interference component
and can be sensitively controlled by the AB phase $\phi$.
In fact, the quantum inference occurs in
electron transport between the two reservoirs.
That is, the electron tunneling from the left (right) reservoir to the
right (left) reservoir is via the interference pathes,
i.e., dot 1 (channel 1) and dot 2 (channel 2).
Apparently, as the aforementioned quantum interference characteristics
are concerned, the cross-correlation noise spectrum
[\Fig{fig2}(b)] appears more prominent than
the auto-correlation counterpart [\Fig{fig2}(a)].
The latter shows the $\phi$--insensitive dips only.
In this sense, one may consider
$S_{\rm LL}(\w)+S_{\rm RR}(\w)$
as the background
to the total circuit noise spectrum $S(\w; \phi)$ of \Eq{Sw}.

The details of $S(\w; \phi)$ depicted in \Fig{fig2}(d) are as follows.
First of all, $S(\w; \phi=0)$, the black--curve in \Fig{fig2}(d),
shows only those non-Markovian quasi-step features,
which rise around $\omega=\pm|\varepsilon_u-\mu_\alpha|$,
as described earlier. Here, $\varepsilon_1=-\varepsilon_2 = 0.35$\,meV
and $\mu_{\rm L}=-\mu_{\rm R}=0.7$\,meV.
There are no Lorentzian-like dips,
which exist in the individual
auto-- and cross--correlation noise spectrums
at $\omega=0$ and $\omega=\pm\Delta\varepsilon$,
but are now completely canceled out.
The non-Markovian peaks at $\omega=-|\varepsilon_u-\mu_\alpha|$
in ${\rm Re}[S_{\rm LR}(\omega;\phi=0)]$,
the black--curve in \Fig{fig2}(b),
are also smeared out in the total circuit noise spectrum.
Interestingly, recovered in $S(\w; \phi\neq 0)$,
the colored curves in \Fig{fig2}(d),
are only those Lorentzian-like dips at $\omega=\pm\Delta\varepsilon$
that specifies the non-degeneracy of the double dots.
Apparently, the depths of the characteristic dips
reach maxima when the AB phase $\phi=\pm\pi$;
see the inset of \Fig{fig2}(d).
It comes from the maximum peak feature of the cross-correlation
noise spectrum ${\rm Re}[S_{\rm LR}(\omega;\phi=\pm\pi)]$
as described above, see \Fig{fig2}(b).
Therefore, the total circuit noise spectrum is also sensitive to the
AB phase.

\begin{figure}
\includegraphics[width=1.0\columnwidth]{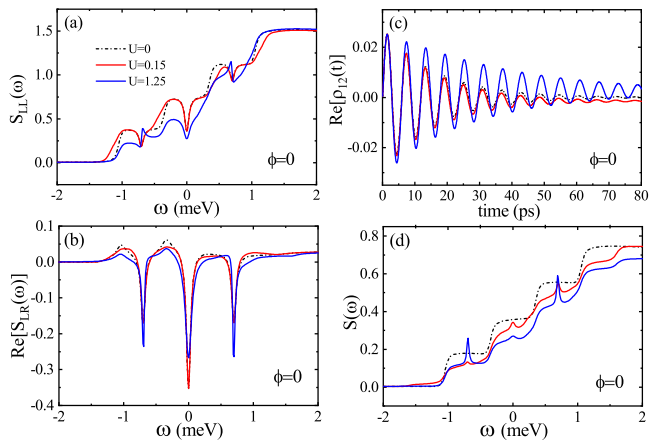}
\caption{ (Color online)
The accurate results for the current noise spectra (in $2\bar I$)
and the real-time dynamics of
${\rm Re}[\rho_{12}(t)]$,
with differenct interdot Coulomb interaction
for AB phase $\phi=0$, indirect interdot coupling parameter $\lambda=1$
and energy splitting $\Delta\varepsilon=0.7{\rm meV}$.
(a) The auto-correlation noise spectrum of the left lead.
 (b) The cross-correlation noise spectrum of the left-right leads.
(c) The time evolution of ${\rm Re}[\rho_{12}(t)]$.
 (d) The circuit noise spectrum.
 The other parameters are the same as in Fig.\,\ref{fig1}.
 }
\label{fig3}
\end{figure}

\subsection{Effect of interdot Coulomb interaction }
\label{thCou}

 Consider now the effect of interdot Coulomb interaction $U$
on the coherent Rabi oscillation dynamics, exemplified
with $\Delta\varepsilon=0.7$\,meV.
Coulomb-assisted transport channels ($\varepsilon_{1,2}+U$)
modifies the characteristic quasi-step at $\omega=\pm|\varepsilon_{1,2}+U-\mu_\alpha|$
and related features in the current noise spectrums.

Figure \ref{fig3} reports the results
of the pristine double--dots system
($\phi=0$), with different values of $U$.
In the weak Coulomb interaction regime ($U=0.15\,{\rm meV}$; red--curves),
where $\mu_{\rm L}>\varepsilon_{1,2}+U,\varepsilon>\mu_{\rm R}$,
all transport channels
are inside the sequential tunneling or dias window.
For the lead-specified noise spectrum components
and the density matrix coherence, \Fig{fig3}(a)--(c),
the Rabi oscillation signals
differ insignificantly from the noninteracting counterparts
(red--curves versus black--curves).
However, the total circuit noise spectrum
$S(\omega)$ as shown in \Fig{fig3}(d) differs distinctly.
It shows the peaks at $\omega=0$ and $\omega=\pm\Delta\varepsilon$,
which would be completely cancelled out if $U=0$.
In other words,
the total circuit noise spectrum is sensitive
to the Coulomb interaction, even when
the Coulomb--assisted transport channels are
inside the bias window.

 In the strong Coulomb interaction regime ($U=1.25\,{\rm meV}$; blue--curves),
with $\varepsilon_{1,2}+U>\mu_{\rm L}>\varepsilon>\mu_{\rm R}$,
where the Coulomb-assisted transport channels
are outside of the bias window.
Akin to the intrinsic Rabi oscillation
via a direct interdot coupling \cite{Luo07085325,Don08033532},
the dips at $\w=\pm\Delta\varepsilon$
in the auto-correlation noise spectrum, $S_{\alpha\alpha}(\w)$,
becomes inflections.
Note that the strong capacitive coupling makes the
double--dots system
in the interdot Coulomb blockade (CB) regime.
This suppresses the zero-frequency shot noise
($S_{\rm LL}(0)=S_{\rm RR}(0)=-{\rm Re}[S_{\rm LR}(0)]$).
Counter-intuitively, as inferred from \Fig{fig3}(b) and (c) the blue--curves,
the strong capacitive coupling gives rise to the
\emph{CB-assisted Rabi interference}.
It enhances the Rabi resonance at $\omega=\pm\Delta\varepsilon$
in the cross-correlation noise spectrum
and the corresponding amplitude of the
density matrix coherence oscillation away from the short-time scale,
i.e., $t>2/\Delta\varepsilon$ as inferred from \Fig{fig3}(c).
The CB--assisted Rabi interference arises from
 the Coulomb-assisted transport channels $\varepsilon_{1,2}+U$
which interfere with
the single-electron impurity channels.
Despite that $\varepsilon_{1,2}+U$ is above the bias window, it
is accessible by finite-frequency current noise spectrum.
The resultant total circuit noise spectrum,
$S(\w)$ in \Fig{fig3}(d),
the blue--curve in comparing to the red--curve,
shows a stronger peak at $\w=\pm\Delta\varepsilon$
but a weaker one at $\w=0$.

\subsection{Coulomb--blockage--assisted Rabi interference
with AB interferometer}
\label{thCouRabi}

\begin{figure}
\includegraphics[width=1.0\columnwidth]{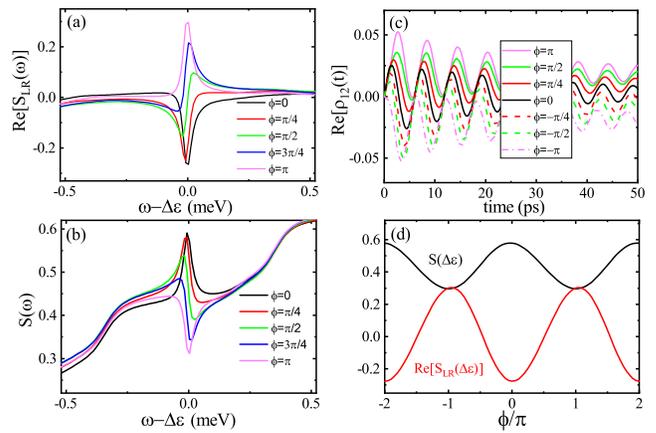}
\caption{ (Color online)
The accurate results for the current noise spectra (in $2\bar I$)
and the real-time dynamics of
${\rm Re}[\rho_{12}(t)]$,
with different AB phase $\phi$ for strong
interdot Coulomb interaction $U=1.25 {\rm meV}$,
indirect coupling parameter $\lambda=1$,
and energy splitting $\Delta\varepsilon=0.7{\rm meV}$.
 (a) The cross-correlation noise spectrum. 
  (b) The circuit noise spectrum.
(c) The time evolution of ${\rm Re}[\rho_{12}(t)]$.
(d) The oscillation signal at $\omega=\Delta\varepsilon$ in
both the cross-correlation noise and circuit noise as a function of AB phase.
 The other parameters are the same as in Fig.\,\ref{fig1}. }
\label{fig4}
\end{figure}

\subsubsection{Emergence of Fano interference}

 Figure 4 depicts the effects of AB phase $\phi$
on those CB--assisted Rabi interference
relevant quantities.
Here, we only plot the absorption noise with $\omega>0$, 
since the Rabi oscillation signals in both the emission and absorption
noise parts are quite similar (see the above firgures).
We also do not repeat the auto-correlation noise spectrum
because it is AB phase $\phi$--insensitive and remains largely like \Fig{fig3}(a).
The fact that the cross-correlation noise is $\phi$--sensitive
but the auto-correlation is not, is common for both noninteracting and interacting scenarios.
The Rabi oscillation, the evolution of the reduced density
matrix off-diagonal element as seen in \Fig{fig4}(c),
is also $\phi$-sensitive.
There is a remarkably distinct feature of CB-assisted Rabi interference with
AB interferometer:
In the total circuit noise spectrum,
the Rabi signatures are peak and dip
at AB phase $\phi=0$ and $\pi$, respectively,
but Fano profiles in between.
Evidently, this remarkable feature arises from
 the cross-correlation noise spectrum, as shown in
  \Fig{fig4}(a) versus (b).
This novel phenomenon could be understood below.

\subsubsection{Understanding via canonical transformation}
We restore to the canonical transformation on the electron operators
in the dots which is given by \cite{Jin18043043},
\be\label{unit}
  \ti d_{1} = ({\hat a}_1+{\hat a}_2)/{\sqrt{2}}\ \ \text{and}
\ \  \ti d_{2} = ({\hat a}_1-{\hat a}_2)/{\sqrt{2}}.
\ee
The two single-electron impurity states become
 $|\ti 1\ra$ and $|\ti 2\ra$, with $|\ti u\ra \equiv \ti d^{\dg}_u|0\ra$.
The Hamiltonians described in \Eqs{Hs}
and \Eq{Hsb} can be rewritten, respectively, as
\be\label{Hstrans}
 \wti H_{\tS} =\frac{\varepsilon_0}{2}\sum_{u=1,2}
  \ti d^{\dg}_{u}\ti d_{u}
 +\frac{\Delta\varepsilon}{2}  (\ti d^{\dg}_{1}\ti d_{2}
  +\ti d^{\dg}_{2}\ti d_{1})
 +U \ti n_1\ti n_2,
\ee
with 
$\ti n_u=\ti d^{\dg}_{u}\ti d_{u}$,
 and
\be\label{Hsb2}
 {\wti H}_{\SB}=\sum_{\alpha u k}
 \left( {\ti t}_{\alpha u k}  {\ti d}^{\dg}_{u}\hat c_{\alpha k}
    +{\rm H.c.}\right),
\ee
with the tunneling coefficients
${\ti t}_{\alpha 1 k} =\sqrt{2}t_{\alpha k}\cos(\phi/4)$
and ${\ti t}_{{\rm L}2 k}={\ti t}^\ast_{{\rm R}2 k}=i\sqrt{2} t_{\alpha k}\sin(\phi/4)$.
They lead to the coupling strengths in
the hybridization spectral function of \Eq{J_Drude}
as \cite{Jin18043043},
 \be\label{tigamma}
 \begin{split}
\wti\Gamma_{\alpha 11}&=2\Gamma\cos^2(\phi/4),~~
\wti\Gamma_{\alpha 22}=2\Gamma\sin^2(\phi/4),
\\
\wti\Gamma_{{\rm L}12}&=\wti\Gamma^\ast_{{\rm L}21}=
\wti\Gamma_{{\rm R}21}=\wti\Gamma^\ast_{{\rm R}12}=i\Gamma\sin(\phi/2).
\end{split}
\ee
Note that here, we considered the symmetrical coupling strength and
$\lambda_{\rm L}=\lambda_{\rm R}=\lambda=1$.
%


%

The observed CB-assisted interference with AB interferometer 
could be understood
based on \Eqs{Hstrans}--(\ref{tigamma}).
First of all, the auto-correlation noise describes the
fluctuations of the electrons tunneling forth and back
from the dots.
Its spectrum is insensitive to the AB phase as
we elaborated earlier, cf.\,\Fig{fig3}(a).
Therefore, as the Rabi characteristic in relation to the
AB phase is concerned, the circuit noise spectrum [\Fig{fig4}(b)] is
effectively opposite to that of cross-correlation [\Fig{fig4}(a)].
The latter will be the focus below.
We identity the following
three cases according to the value of AB phase.

 (\emph{i}) For $\phi=2m\pi$ [black-curve in \Fig{fig4}(a)], we have
 either $\wti\Gamma_{\alpha 11}=0$ or $\wti\Gamma_{\alpha 22}=0$,
with odd or even $m$, respectively, and accordingly $\wti\Gamma_{\alpha 12}=0$.
The electron tunnels through only one of the
transformed single--electron channels,
either $|\ti 1\ra$ or $|\ti 2\ra$.
However, the second term in \Eq{Hstrans}
 induces the transition between
 $|\ti 1\ra$ and $|\ti 2\ra$.
The interdot Coulomb interaction thus plays the role.
Current transports through DD-AB system via the interference between
Coulomb-assisted and the single-electron
impurity channels. The underlying Rabi resonance gets enhanced
and so do for the cross-correlation noise spectrums.
The Rabi signature is a conventional dip,
since only one single--electron channel is allowed.

(\emph{ii})
 For $\phi=(2m+1)\pi$ [purple-curve in \Fig{fig4}(a)], we have
$\wti\Gamma_{\alpha 11}=\wti\Gamma_{\alpha 22}=|\wti\Gamma_{\alpha 12}|=\Gamma$.
This is the case of a full interference between both single-electron
channels with equal probability.
The Rabi signature is a peak, opposite to the one channel case (\emph{i}).

 (\emph{iii}) For $\phi\neq n\pi$
in between cases (\emph{i}) and (\emph{ii}), both single-electron
channels are involved, but with different probabilities.
The large one is the fast channel and vice versa \cite{Jin18043043,Saf03136801}.
This situation is actually the dynamical Coulomb blockade
regime \cite{Jin18043043,Saf03136801}.
Current transports through DD-AB system via the interference between
both flux-dependence
fast and slow channels.
That induces the flux-dependence Rabi--Fano profiles
in the cross-correlation noise spectrum.

\section{Summary}
\label{thsum}

In summary, we have thoroughly investigated the
nonequilibirum quantum noise spectrum of the transport current through
a nondegenerate double--dot embedded in an AB interferometer.
Based on accurate DEOM evaluations,
we demonstrate the rich spectroscopic signatures
in relation to the Rabi oscillation between the two charge states in the double dots.
For noninteracting double dots, the total circuit noise spectrum
only shows non-Markovian quasi-steps at $\phi=0$,
but Rabi dip at nonzero AB phase.
For strong interdot Coulomb interaction,
the spectrum signatures are peak and dip
at $\phi=0$ and $\phi=\pi$, respectively,
but Fano profiles at other values of AB phase in between.
Occurs there a remarkable
Coulomb-blockade-assisted Rabi interference,
arising from the interplay between various mechanisms.
These include the environment-assisted indirect interdot coupling,
coherent Rabi oscillation, lead-specified current-current interference,
interdot Coulomb interaction, and further AB phase.
All these mechanisms and the interplay between them
are elaborated in the present work.
The rich characteristics are inaccessible
in the average current and zero-frequency shot noise.
  To the best of our knowledge, these results are the first
uncovered in the present work and may have potential
applications in the field of quantum computer and quantum information.

\acknowledgments
The author gratefully thanks Professor YiJing Yan for insightful discussions.
Support from the Natural Science Foundation
of China (No.\ 4511447006) and the NKRDP of China (2016YFA0301802)
is acknowledged.


\appendix
\section{The DEOM theory}
\label{thDEOM}

 In this appendix, we briefly outline the derivation of
 the two-time current-current correlation function based on the
 DEOM approach. The details see the references \cite{Yan14054105,Jin15234108,Yan16110306}.
 Consider an electron transport setup,
in which an impurity system ($H_{\tS}$) is sandwiched by electrodes bath ($h_{\B}$),
under electric bias potential ($eV=\mu_{{\rm L}}-\mu_{{\rm R}}$)
applied cross the leads, $\alpha = {\rm L}$ and R.
The total Hamiltonian reads $H_{\T}=H_{\tS}+h_{\B}+H_{\SB}$.
The system Hamiltonian $H_{\tS}$ includes the electron-electron
interaction and is given in terms of local electron creation $\hat a^{\dg}_{u}$
(annihilation $\hat a_{u}$) operators of the specified system, for instance,
see \Eq{Hs} in the present study.
The electrodes bath is modeled as noninteracting electrons
reservoir described by \Eq{HB}. The transport coupling between
the reservoir and the system is given by \Eq{Hsb} which is rewritten
as
\be\label{Hsb1}
  H_{\SB}\!=\!\sum_{\alpha u }\left(\hat a^{\dg}_{u}  \hat F_{\alpha u}
    +{\rm H.c.} \right),
\ee
with $\hat F_{\alpha u}=\sum_k e^{i\phi_{\alpha u}}t_{\alpha u k} c_{\alpha k}$.
For the calculation of the current noise spectrum, we should restore to
the technique of the dissipaton decomposition on the
hybrid bath \cite{Yan14054105,Jin15234108,Yan16110306}, i.e.,
$\hat  F^{\sigma}_{\alpha u}
    \equiv \sum_{k=1}^{K} \bar\sigma\hat f^{\sigma}_{\alpha u k} $ in \Eq{Hsb1},
with the dissipatons operator $ f^{\sigma}_{\alpha u k} $ satisfying $\big\la\hat f^{\sigma}_{\alpha u k}(t)\hat f^{\sigma'}_{\alpha' v k'}(0)\big\ra_{\B}
 =-\delta_{\bar\sigma\sigma'}\delta_{\alpha\alpha'}\delta_{kk'}\,
   \eta^{\sigma}_{\alpha uv k} e^{-\gamma^{\sigma}_{\alpha k} t}$.
This dissipaton decomposition arises from the
\emph{nonequilibrium} interacting reservoirs bath correlation
functions \cite{Jin08234703,Zhe121129,Li12266403,Ye16608,Hu10101106,Hu11244106},
in an exponent expansion form of
\be\label{Csigma_exp}
\big\la \hat F^{\sigma}_{\alpha u}(t)
 \hat F^{\bar\sigma}_{\alpha v}(0)\big\ra_{\B}
= \sum_{\kappa=1}^{K}
  \eta^{\sigma}_{\alpha uv\kappa}e^{-\gamma^{\sigma}_{\alpha\kappa}t}.
\ee
Such an exponential expansion in \Eq{Csigma_exp} is realized
via a sum-overpoles
decomposition for the Fourier integrand of the relation
which is the so-called fluctuation-dissipation theorem
\cite{Jin08234703,Zhe121129,Kle09,Wei08,Yan05187}:
 $\big\la \hat F^{\sigma}_{\alpha u}(t)
 \hat F^{\bar\sigma}_{\alpha v}(0)\big\ra_{\B}
=\frac{1}{\pi}\int\,d\omega\,e^{\sigma i(\omega+\mu_{\alpha}\!)t}
   \frac{J^{\sigma}_{\alpha uv}(\omega)}
    {1+e^{\sigma\beta\omega}}$,
followed by Cauchy's contour integration.
Here, we adopt the Lorentz form of the
the hybridization function, see \Eq{J_Drude}, with
$J^{+}_{\alpha vu}(\omega)=J^{-}_{\alpha uv}(\omega)=J_{\alpha uv}(\omega)$.
 The exponents $\gamma^{\sigma}_{\alpha\kappa}$ in \Eq{Csigma_exp}
 arising from both the Fermi function and the hybridization
function.

 For bookkeeping, we adopt the abbreviations,
 $j\equiv(\sigma\alpha u\kappa)$ and $\bar j\equiv(\bar\sigma\alpha u \kappa)$,
for the collective indexes in fermionic dissipatons, such
that $f_j\equiv f^{\sigma}_{\alpha ku}$ and so on.
The superindex $\sigma =+,-$ (and $\bar\sigma$ is its opposite sign)
is used to redefine the
fermion creation and annihilation operators, e.g.,
$\hat a^+_{u}\equiv\hat  a^\dg_{u}$
and $\hat a^-_{u}\equiv \hat a_{u}$ in \Eq{calAC}.
The quantum coherence and/or decoherence dynamics of the impurity systems
 are described by the reduced density matrix, $\rho(t)\equiv {\rm tr}_{\B}\rho_{\rm tot}(t)$, i.e.,
the partial trace of the total density operator $\rho_{\rm tot}(t)$ over the electrode bath degrees of freedom.
Dynamical variables in DEOM are the reduced dissipaton density
operators (DDOs),
\be\label{DDO_def}
 \rho^{(n)}_{\bf j}(t)\equiv \rho^{(n)}_{j_1\cdots j_n}(t)\equiv
 {\rm tr}_{\B}\Big[\big(\hat f_{j_n}\cdots\hat f_{j_1}\big)^{\circ}
  \rho_{\rm tot}(t)\Big]\, .
\ee
Here, $\big(\hat f_{j_n}\!\cdots\!\hat f_{j_1}\big)^{\circ}$
specifies an \emph{ordered} set of $n$ \emph{irreducible} dissipatons.
A swap of any two irreducible fermionic dissipatons
causes a minus sign, such that
\be\label{irr_fermion}
 \big(\hat f_{j}\hat f_{j'}\big)^{\circ}=-\big(\hat f_{j'}\hat f_{j}\big)^{\circ}.
\ee
From \Eq{DDO_def}, it is clear that the reduced system density operator is just
$\rho(t) \equiv \rho^{(0)}(t)$.
The DEOM formalism addresses also
the hybridizing bath subspace dynamics.
After some algorihm together with {\it Wick's theorem},
the DEOM formalism has been obtained as \cite{Yan14054105,Jin15234108,Yan16110306}
\begin{align}\label{DEOM}
  \dot\rho^{(n)}_{\bf j}&=-\bigg(i{\cal L}_{\tS}
  +\sum_{r=1}^n \gamma_{j_r}\bigg)\rho^{(n)}_{\bf j}
  -i\sum_{j} {\cal A}_{\bar j}\rho^{(n+1)}_{{\bf j}j}    \nl
&\quad
  -i \sum_{r=1}^n (-)^{n-r}{\cal C}_{j_r}\rho^{(n-1)}_{{\bf j}^-_r}.
\end{align}
The Grassmannian superoperators, ${\cal A}_{\bar j}\equiv {\cal A}^{\bar\sigma}_{\alpha u\kappa} = {\cal A}^{\bar\sigma}_{u}$
and ${\cal C}_{j}\equiv {\cal C}^{\sigma}_{\alpha u\kappa}$ in \Eq{DEOM},
 are defined via\cite{Jin08234703,Li12266403,Ye16608}
\be\label{calAC}
\begin{split}
 {\cal A}^{\sigma}_{u} \Opm &\equiv
    a^{\sigma}_{ u}\Opm \pm \Opm \hat a^{\sigma}_{u}
 \equiv \big[\hat  a^{\sigma}_{u},\Opm\big]_\pm \, ,
\\
 {\cal C}^{\sigma}_{\alpha u\kappa} \Opm  &\equiv
  \sum_{v} \big(\eta^{\sigma}_{\alpha uv\kappa}\hat  a^{\sigma}_{v}\Opm
  \mp \eta^{\bar \sigma\,{\ast}}_{\alpha uv\kappa}\Opm \hat a^{\sigma}_{\kappa}\big) .
\end{split}
\ee
Here, $\Opm$ is an arbitrary operator,
with even ($+$) or odd ($-$)  fermionic parity,
such as $\rho^{(2m)}$ or $\rho^{(2m+1)}$, respectively.

The hierarchical structure in \Eq{DEOM} is formally identical to
that derived originally via the calculus on
path integral influential functions \cite{Jin08234703}.
 All $\big\{\rho^{(n)}_{\bf j}\big\}$
are now the physically well-defined DDOs, with \Eq{DDO_def}
that goes by the mathematical irreducibility.
The hierarchical coupling can be simply
truncated by setting all $\rho^{(n>L)}_{\bf j}=0$,
at a sufficiently large $L$.
While all $L$--body dissipatons dynamics are treated exactly,
the resulting closed DEOM for $\big\{\rho^{(n)}_{\bf j}; n=0,1,\cdots,L\big\}$
represent also a dynamical mean--field scheme
for higher--order DDOs. In this sense,
DEOM is naturally a nonperturbative many-particle theory.
  The present algebraic construction, based on the quasi-particle
(dissipaton) description of hybridizing bath, renders
the DEOM, represented by \Eq{DEOM}, a correlated system and bath
dynamics theory. It can be used in the accurate
evaluation on nonequilibrium properties of
current noise spectrum below.

From the definition, $ \hat I_{\alpha}\equiv -\frac{d}{dt}
   \big(\sum_{k}c^{\dg}_{\alpha k}c_{\alpha k}\big)$,
 the current operator, for the electron transfer from $\alpha$-reservoir to
 the impurity system, reads
\be\label{curr_operator}
 \hat I_{\alpha}
 =-i\sum_\mu \big(\hat  a^{\dg}_u \hat F_{\alpha u}
   -\hat F^{\dg}_{\alpha u}\hat  a _u \big)
   =  -i \sum_{j_{\alpha}\in j}\ti a_{\bar j}\hat f_{j} \,.
\ee
The second identity in \Eq{curr_operator} is expressed in
the dissipatons decomposition, where
 $\ti a_{\bar j}\equiv \ti a^{\bar\sigma}_{\alpha u k}=\ti a^{\bar\sigma}_{u}
 \equiv \bar \sigma \hat a^{\bar\sigma}_{u}$
and $j_{\alpha}\equiv \{ \sigma u k\}\in j\equiv\{\sigma\alpha u k\}$.
The mean current can then be evaluated in terms of the first-tier DDOs as
\be \label{curr}
  I_{\alpha}(t)
= {\rm Tr}_{\T}\big[\hat I_{\alpha}\rho_{\rm tot}(t)\big] =
  -i \sum_{j_{\alpha}\in j} {\rm tr}_{\tS}\!\big[\ti a_{\bar j}\rho^{(1)}_{j}(t)\big].
\ee
The trace ${\rm tr}_{\tS}$ runs over the system degrees of freedom.
It has been used in evaluating both steady-state and time-dependent transient
current in various situations\cite{Zhe08184112,Zhe08093016,Zhe09164708},
including dynamical Kondo memory \cite{Zhe13086601}
and thermopower \cite{Ye14165116} responses.

Similarly, the two-time current-current correlation function can be evaluated as
\begin{align}\label{CorrI}
 \La\hat I_{\alpha}(t)\hat I_{\alpha'}(0)\Ra 
&= {\rm Tr}_{\T}\big[\hat I_{\alpha}\rho_{\rm tot}(t;\alpha')\big]
\nl&= - i\sum_{j_{\alpha}\in j}
 {\rm tr}_{\tS} \big[\ti a_{\bar j}\, {\rho}^{(1)}_j(t;\alpha')\big],
\end{align}
where
\be\label{rhoT_alpha}
 \rho_{\rm tot}(t;\alpha')\equiv e^{-i{\cal L}_{\T}t}
 \big(\hat I_{\alpha'}\rho^{\rm st}_{\rm tot}\big),
\ee
with  $\rho_{\T}(t=0;\alpha')=\hat I_{\alpha'}\rho^{\rm st}_{\T}$
and $\rho^{\rm st}_{\T}$ denoting
the steady-state total composite density operator, under the bias voltage
of $V=\mu_{\rm L}-\mu_{\rm R}$.
 The key step for the calculation of the correlation function based on
 the DEOM evaluation is to identify the initial DDOs,
$\big\{{\rho}^{(n)}_{\bm j}(0; \alpha')\big\}$ that are associated with
$\rho_{\T}(0;\alpha')=\hat I_{\alpha'}\rho^{\rm st}_{\T}$.
Based on the underlying dissipaton algebra and the
generalized Wick's theorem (for the details see Refs.\onlinecite{Yan14054105,Jin15234108,Yan16110306}).
The initial values of DDOs in \Eq{CorrI} are given by
\begin{align}\label{initial_final}
 \rho^{(n)}_{\bm j}(0; \alpha')
&\!=\! -i\!\!\!\sum_{j'_{\alpha'}\in j'}{\ti a}_{\bar j'} \rho^{(n+1);{\rm st}}_{{\bm j}j'}
\!\! -\!\! i\!\sum_{r=1}^n(-)^{n-r} \wti C_{j_r}\rho^{(n-1);{\rm st}}_{{\bm j}^{-}_r},
\end{align}
where, $\big\{\rho^{(n\pm 1);{\rm st}}\big\}$ are
the steady-state solutions  to DEOM (\ref{DEOM})
by using the conditions, $\{\dot\rho^{(n);{\rm st}}_{\bf j}=0; \forall n\}$
together with the normalization constraint, ${\rm tr}\rho^{(0)}=1$,
 under given constant bias potential.
The resulting initial $\big\{\rho^{(n)}_{\bm j}(0; \alpha')\big\}$
are then propagated, by using the real-time dynamics of \Eq{DEOM} again,
to obtain $\big\{\rho^{(n)}_{\bm j}(t; \alpha')\big\}$.
Finally, the lead-specified current correlation function,
$\La\hat I_{\alpha}(t)\hat I_{\alpha'}(0)\Ra$, is evaluated
according to \Eq{CorrI}. Consequently,
we can get the current noise spectrum \Eq{Sw_alp} together with
\Eq{Cw_alp}.

\begin{figure}
\includegraphics[width=1.0\columnwidth]{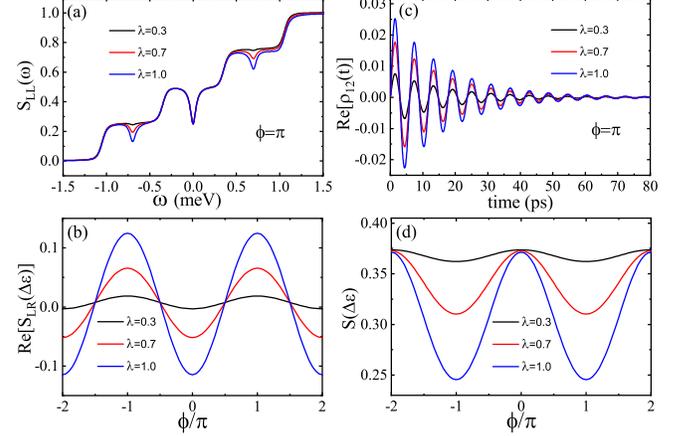}
\caption{ (Color online)
The accurate results for the current noise spectra (in $2\bar I$)
and the real-time dynamics of
the reduced density matrix off-diagonal element,
with different indirect coupling parameter $\lambda$
for $U=0$ and $\Delta\varepsilon=0.7{\rm meV}$.
(a) The frequency-dependent auto-correlation current noise
for AB phase $\phi=\pi$.
(b) The cross-correlation noise spectrum at the oscillation
signal in the absorption part as a function of
the AB phase.
(c) The time evolution of the coherence term ${\rm Re}[\rho_{12}(t)]$
 for AB phase $\phi=\pi$.
(d) The total circuit noise spectrum at the Rabi
signal in the absorption part as a function of
the AB phase.
The other parameters are the same as in Fig.\,\ref{fig1}.
}
\label{fig5}
\end{figure}

\section{The effect of the parameter $\lambda$}  
\label{thlam}


 As expected, both the Rabi signal in the noise spectrum
and the Rabi oscillation dynamics
become weaker as the indirect coupling parameter decreases.
Figure \ref{fig3}(a) and (c) report this characteristic,
in the auto-correlation noise spectrum and
the reduced density matrix coherence evolution,
respectively, as demonstrated with $\phi=\pi$.
The oscillatory quantum interference pattern
disappears if there is no indirect coupling ($\lambda=0$).
Meanwhile, the effect of AB phase $\phi$  on
the noise spectrum remain qualitatively the same
as \Fig{fig2}(a) and (b).
It affects little on  the auto-correlation noise spectrum,
but dramatically on the cross-correlation one.
The latter dictates the effect of AB phase $\phi$
on the total circuit noise spectrum, $S(\w)$ of \Eq{Sw}.

 Figure \ref{fig3}(b) and (d) depict the characteristic
peak/dip values of
the cross-correlation and circuit noise spectrums, respectively,
at the Rabi frequency $\w=\Delta\varepsilon$.
The observations here can be understood via
$\Gamma_{\alpha12}$ of \Eq{Gam}.
This cross--type of system--reservoirs coupling
is responsible for the Rabi interference
in transport.
It is nonzero only when the indirect interdot tunnel coupling parameter
$\lambda\neq 0$.
Its absence leads to the electrons tunneling through
channels 1 and 2 being independent.
The indirect interdot tunnel coupling gives
rise to the indistinguishability between
the two channels. This leads further to the coherence
between the two charge states ($|1\ra$ and $|2\ra$) in the double-dot,
and the Rabi oscillation occurs
whenever $|1\ra$ and $|2\ra$ are also nondegenerate.


\begin{thebibliography}{10}

\bibitem{Sch97417}
R.~Schuster et~al.,
\newblock Nature {\bf 385}, 417 (1997).

\bibitem{Hol01256802}
A.~W. Holleitner, C.~R. Decker, H.~Qin, K.~Eberl, and R.~H. Blick,
\newblock Phys. Rev. Lett. {\bf 87}, 256802 (2001).

\bibitem{Sig06036804}
M.~Sigrist et~al.,
\newblock Phys. Rev. Lett. {\bf 96}, 036804 (2006).

\bibitem{Sig04066802}
M.~Sigrist et~al.,
\newblock Phys. Rev. Lett. {\bf 93}, 066802 (2004).

\bibitem{Kub06205310}
T.~Kubo, Y.~Tokura, T.~Hatano, and S.~Tarucha,
\newblock Phys. Rev. B {\bf 74}, 205310 (2006).

\bibitem{Hat11076801}
T.~Hatano et~al.,
\newblock Phys. Rev. Lett. {\bf 106}, 076801 (2011).

\bibitem{Liu16045403}
J.-H. Liu, M.~W.-Y. Tu, and W.-M. Zhang,
\newblock Phys. Rev. B {\bf 94}, 045403 (2016).

\bibitem{Kan04117}
K.~Kang and S.~Y. Cho,
\newblock J. Phys.: Condens. Matter {\bf 16}, 117 (2004).

\bibitem{Szt07386224}
D.~Sztenkiel and R.~\'{S}wirkowicz,
\newblock J. Physics: Condensed Matter {\bf 19}, 386224 (2007).

\bibitem{Kon013855}
J.~{K\"{o}nig} and Y.~Gefen,
\newblock Phys. Rev. Lett. {\bf 86}, 3855 (2001).

\bibitem{Kon02045316}
J.~K\"onig and Y.~Gefen,
\newblock Phys. Rev. B {\bf 65}, 045316 (2002).

\bibitem{Li09521}
F.~Li, H.~J. Jiao, H.~Wang, J.~Y. Luo, and X.~Q. Li,
\newblock Physica E: Low-dimensional Systems and Nanostructures {\bf 41}, 521
  (2009).

\bibitem{Tok07113}
Y.~Tokura, H.~Nakano, and T.~Kubo,
\newblock New J. Phys. {\bf 9}, 113 (2007).

\bibitem{Bed14235411}
S.~Bedkihal and D.~Segal,
\newblock Phys. Rev. B {\bf 90}, 235411 (2014).

\bibitem{Rep16165425}
E.~V. Repin and I.~S. Burmistrov,
\newblock Phys. Rev. B {\bf 93}, 165425 (2016).

\bibitem{Bur992070}
G.~Burkard, D.~Loss, and D.~P. DiVincenzo,
\newblock Phys. Rev. B {\bf 59}, 2070 (1999).

\bibitem{Har13235426}
R.~H{\"a}rtle, G.~Cohen, D.~Reichman, and A.~Millis,
\newblock Phys. Rev. B {\bf 88}, 235426 (2013).

\bibitem{Bed13045418}
S.~Bedkihal, M.~Bandyopadhyay, and D.~Segal,
\newblock Phys. Rev. B {\bf 87}, 045418 (2013).

\bibitem{Tu12115453}
M.~W.-Y. Tu, W.-M. Zhang, J.~S. Jin, O.~Entin-Wohlman, and A.~Aharony,
\newblock Phys. Rev. B {\bf 86}, 115453 (2012).

\bibitem{Tu11115318}
M.~W.-Y. Tu, W.-M. Zhang, and J.~S. Jin,
\newblock Phys. Rev. B {\bf 83}, 115318 (2011).

\bibitem{Bed12155324}
S.~Bedkihal and D.~Segal,
\newblock Phys. Rev. B {\bf 85}, 155324 (2012).

\bibitem{Jin18043043}


\bibitem{Lei09156803}
M.~Leijnse, M.~R. Wegewijs, and M.~H. Hettler,
\newblock Phys. Rev. Lett. {\bf 103}, 156803 (2009).

\bibitem{Her927061}
S.~Hershfield,
\newblock Phys. Rev. B {\bf 46}, 7061 (1992).

\bibitem{Bla001}
Y.~M. Blanter and M.~B\"{u}ttiker,
\newblock Phys. Rep. {\bf 336}, 1 (2000).

\bibitem{Imr02}
I.~Imry,
\newblock {\em Introduction to Mesoscopic Physics},
\newblock Oxford university press, 2002.

\bibitem{Naz03}
{\em Quantum Noise in Mesoscopic Physics},
\newblock Kluwer, Dordrecht, 2003,
\newblock edited by Y. V. Nazarov.

\bibitem{Pic97162}
R.~de~Picciotto et~al.,
\newblock Nature {\bf 389}, 162 (1997).

\bibitem{Rez99238}
M.~Reznikov, R.~d. Picciotto, T.~G. Griffiths, M.~Heiblum, and V.~Umansky,
\newblock Nature {\bf 399}, 238 (1999).

\bibitem{Bid09236802}
A.~Bid, N.~Ofek, M.~Heiblum, V.~Umansky, and D.~Mahalu,
\newblock Phys. Rev. Lett. {\bf 103}, 236802 (2009).

\bibitem{Koz003398}
A.~A. Kozhevnikov, R.~J. Schoelkopf, and D.~E. Prober,
\newblock Phys. Rev. Lett. {\bf 84}, 3398 (2000).

\bibitem{Lef03067002}
F.~Lefloch, C.~Hoffmann, M.~Sanquer, and D.~Quirion,
\newblock Phys. Rev. Lett. {\bf 90}, 067002 (2003).

\bibitem{Zha06085106}
G.~B. Zhang, S.~J. Wang, and L.~Li,
\newblock Phys. Rev. B {\bf 74}, 085106 (2006).

\bibitem{Fan07205312}
T.-F. Fang, S.-J. Wang, and W.~Zuo,
\newblock Phys. Rev. B {\bf 76}, 205312 (2007).

\bibitem{Bre11155305}
D.~Breyel and A.~Komnik,
\newblock Phys. Rev. B {\bf 84}, 155305 (2011).

\bibitem{Agu001986}
R.~Aguado and L.~P. Kouwenhoven,
\newblock Phys. Rev. Lett. {\bf 84}, 1986 (2000).

\bibitem{Deb03203}
R.~Deblock, E.~Onac, L.~Gurevich, and L.~P. Kouwenhoven,
\newblock Science {\bf 301}, 203 (2003).

\bibitem{Ona06176601}
E.~Onac et~al.,
\newblock Phys. Rev. Lett. {\bf 96}, 176601 (2006).

\bibitem{Zak07236803}
E.~Zakka-Bajjani et~al.,
\newblock Phys. Rev. Lett. {\bf 99}, 236803 (2007).

\bibitem{Bas10166801}
J.~Basset, H.~Bouchiat, and R.~Deblock,
\newblock Phys. Rev. Lett. {\bf 105}, 166801 (2010).

\bibitem{Bas12046802}
J.~Basset et~al.,
\newblock Phys. Rev. Lett. {\bf 108}, 046802 (2012).

\bibitem{Del18041412}
R.~Delagrange, J.~Basset, H.~Bouchiat, and R.~Deblock,
\newblock Phys. Rev. B {\bf 97}, 041412 (2018).

\bibitem{Ent07193308}
O.~Entin-Wohlman, Y.~Imry, S.~A. Gurvitz, and A.~Aharony,
\newblock Phys. Rev. B {\bf 75}, 193308 (2007).

\bibitem{Li05066803}
X.~Q. Li, P.~Cui, and Y.~J. Yan,
\newblock Phys. Rev. Lett. {\bf 94}, 066803 (2005).

\bibitem{Bar06017405}
S.~D. Barrett and T.~M. Stace,
\newblock Phys. Rev. Lett. {\bf 96}, 017405 (2006).

\bibitem{Gab08026601}
J.~Gabelli and B.~Reulet,
\newblock Phys. Rev. Lett. {\bf 100}, 026601 (2008).

\bibitem{Wab09016802}
J.~Wabnig, B.~W. Lovett, J.~H. Jefferson, and G.~A.~D. Briggs,
\newblock Phys. Rev. Lett. {\bf 102}, 016802 (2009).

\bibitem{Eng04136602}
H.-A. Engel and D.~Loss,
\newblock Phys. Rev. Lett. {\bf 93}, 136602 (2004).

\bibitem{Jin11053704}
J.~S. Jin, X.~Q. Li, M.~Luo, and Y.~J. Yan,
\newblock J. Appl. Phys. {\bf 109}, 053704 (2011).

\bibitem{Jin13025044}
J.~S. Jin, M.~Marthaler, P.-Q. Jin, D.~Golubev, and G.~Sch\"on,
\newblock New J. Phys. {\bf 15}, 025044 (2013).

\bibitem{Rot09075307}
E.~A. Rothstein, O.~Entin-Wohlman, and A.~Aharony,
\newblock Phys. Rev. B {\bf 79}, 075307 (2009).

\bibitem{Wan13035129}
S.~K. Wang, X.~Zheng, J.~S. Jin, and Y.~J. Yan,
\newblock Phys. Rev. B {\bf 88}, 035129 (2013).

\bibitem{Mac62}
D.~K.~C. MacDonald,
\newblock {\em Noise and Fluctuations: An Introduction},
\newblock Wiley, New York, 1962,
\newblock Ch.\ 2.2.1.

\bibitem{Don08033532}
B.~Dong, X.~L. Lei, and N.~J.~M. Horing,
\newblock J. Appl. Phys. {\bf 104}, 033532 (2008).

\bibitem{Ziv11115304}
M.~Zivkovic, B.~W. Langley, I.~Djuric, and C.~P. Search,
\newblock Phys. Rev. B {\bf 83}, 115304 (2011).

\bibitem{Yan14054105}
Y.~J. Yan,
\newblock J. Chem. Phys. {\bf 140}, 054105 (2014).

\bibitem{Yan16110306}
Y.~J. Yan, J.~S. Jin, R.~X. Xu, and X.~Zheng,
\newblock Frontiers Phys. {\bf 11}, 110306 (2016).

\bibitem{Jin15234108}
J.~S. Jin, S.~K. Wang, X.~Zheng, and Y.~J. Yan,
\newblock J. Chem. Phys. {\bf 142}, 234108 (2015).

\bibitem{Agu04206601}
R.~Aguado and T.~Brandes,
\newblock Phys. Rev. Lett. {\bf 92}, 206601 (2004).

\bibitem{Luo07085325}
J.~Y. Luo, X.~Q. Li, and Y.~J. Yan,
\newblock Phys. Rev. B {\bf 76}, 085325 (2007).

\bibitem{Kie07206602}
G.~Kie$\beta$lich, E.~Sch{\"{o}}ll, T.~Brandes, F.~Hohls, and R.~J. Haug,
\newblock Phys. Rev. Lett. {\bf 99}, 206602 (2007).

\bibitem{Tu08235311}
M.~W.~Y. Tu and W.-M. Zhang,
\newblock Phys. Rev. B {\bf 78}, 235311 (2008).

\bibitem{Mac06085324}
J.~Maciejko, J.~Wang, and H.~Guo,
\newblock Phys. Rev. B {\bf 74}, 085324 (2006).

\bibitem{Jin08234703}
J.~S. Jin, X.~Zheng, and Y.~J. Yan,
\newblock J. Chem. Phys. {\bf 128}, 234703 (2008).

\bibitem{Zhe121129}
X.~Zheng et~al.,
\newblock Prog. Chem. {\bf 24}, 1129 (2012),
\newblock http://www.progchem.ac.cn/EN/Y2012/V24/I06/1129.

\bibitem{Li12266403}
Z.~H. Li et~al.,
\newblock Phys. Rev. Lett. {\bf 109}, 266403 (2012).

\bibitem{Zhe13086601}
X.~Zheng, Y.~J. Yan, and M.~Di~Ventra,
\newblock Phys. Rev. Lett. {\bf 111}, 086601 (2013).

\bibitem{Hou15104112}
D.~Hou et~al.,
\newblock J. Chem. Phys. {\bf 142}, 104112 (2015).

\bibitem{Ye16608}
L.~Z. Ye et~al.,
\newblock WIREs Comp. Mol. Sci. {\bf 6}, 608 (2016).

\bibitem{Saf03136801}
S.~S. Safonov et~al.,
\newblock Phys. Rev. Lett. {\bf 91}, 136801 (2003).

\bibitem{Hu10101106}
J.~Hu, R.~X. Xu, and Y.~J. Yan,
\newblock J. Chem. Phys. {\bf 133}, 101106 (2010).

\bibitem{Hu11244106}
J.~Hu, M.~Luo, F.~Jiang, R.~X. Xu, and Y.~J. Yan,
\newblock J. Chem. Phys. {\bf 134}, 244106 (2011).

\bibitem{Kle09}
H.~Kleinert,
\newblock {\em Path Integrals in Quantum Mechanics, Statistics, Polymer
  Physics, and Financial Markets},
\newblock World Scientific, Singapore, 5th edition, 2009.

\bibitem{Wei08}
U.~Weiss,
\newblock {\em Quantum Dissipative Systems},
\newblock World Scientific, Singapore, 2008,
\newblock 3rd ed. Series in Modern Condensed Matter Physics, Vol.\ 13.

\bibitem{Yan05187}
Y.~J. Yan and R.~X. Xu,
\newblock Annu. Rev. Phys. Chem. {\bf 56}, 187 (2005).

\bibitem{Zhe08184112}
X.~Zheng, J.~S. Jin, and Y.~J. Yan,
\newblock J. Chem. Phys. {\bf 129}, 184112 (2008).

\bibitem{Zhe08093016}
X.~Zheng, J.~S. Jin, and Y.~J. Yan,
\newblock New J. Phys. {\bf 10}, 093016 (2008).

\bibitem{Zhe09164708}
X.~Zheng, J.~S. Jin, S.~Welack, M.~Luo, and Y.~J. Yan,
\newblock J. Chem. Phys. {\bf 130}, 164708 (2009).

\bibitem{Ye14165116}
L.~Z. Ye et~al.,
\newblock Phys. Rev. B {\bf 90}, 165116 (2014).

\end{thebibliography}
\end{document}